\documentclass[twocolumn,aps,showpacs,amssymb,floatfix]{revtex4}
\input epsfig.sty
\newcommand{\be}{\begin{equation}}
\newcommand{\ee}{\end{equation}}
\newcommand{\beq}{\begin{eqnarray}}
\newcommand{\eeq}{\end{eqnarray}}
\usepackage{bm}
\usepackage{graphicx}%

\begin{document}
\setcounter{figure}{\arabic{figure}}

\title{The N to $\Delta$ 
electromagnetic transition form factors from Lattice QCD}
\author{C.~Alexandrou~$^a$,
Ph.\ de Forcrand~$^b$, H.~Neff~$^c$,
 J.~W.~Negele~$^d$, W. Schroers~$^d$ and A. Tsapalis~$^a$}
\affiliation{{$^a$ Department of Physics, University of Cyprus, CY-1678 Nicosia, Cyprus}\\{$^b$ ETH-Z\"urich, CH-8093 Z\"urich and CERN Theory Division, CH-1211 Geneva 23, Switzerland}\\
{$^c$
 Physics Department, Boston University,  Massachusetts 02215, U.S.A.}\\
{$^d$ 
Center for Theoretical Physics, Laboratory for
Nuclear Science and Department of Physics, Massachusetts Institute of
Technology, Cambridge, Massachusetts 02139, U.S.A.}}

\date{\today}%

\begin{abstract}
The
 magnetic dipole, the electric quadrupole and the Coulomb quadrupole
 amplitudes  for the transition $\gamma N\rightarrow \Delta$
are calculated in quenched lattice QCD at $\beta=6.0$
with Wilson fermions. 
Using a new method combining an optimal combination of  interpolating fields for the $\Delta$ and an overconstrained analysis, we  obtain statistically accurate 
results for the dipole form factor
and for the ratios   of the electric and Coulomb   quadrupole 
amplitudes
to the  magnetic dipole amplitude, $R_{EM}$ and  $ R_{SM}$, 
up to momentum transfer squared  1.5~GeV$^2$. We show for the first time using lattice QCD that both  $R_{EM}$ and  $ R_{SM}$ are non-zero and negative, in qualitative agreement with experiment and indicating the presence of deformation in the N- $\Delta$ system.

\end{abstract}

 
\maketitle
 
 

Deformation is an important and well studied phenomenon in atomic and 
nuclear physics, and it is desirable to understand whether it also arises 
in low-lying hadrons and if so, why.
 For classical and quantum systems with spins larger  than $1/2$,  the
 one-body quadrupole operator provides a convenient characterization of 
deformation.
 Experimentally, however, in the excited spin 3/2 $\Delta$, which can have 
a non-zero quadrupole moment, it is not practical to measure it, and the spin 
1/2 nucleon, which is easily accessible to measurement, cannot have a 
spectroscopic quadrupole moment.  Hence, the experiment of choice to reveal 
the presence of deformation in the low-lying baryons is measuring the
 N -$\Delta$ transition amplitude, and significant effort has been 
devoted to photoproduction 
 experiments on the nucleon at Bates~\cite{Bates} and Jefferson Lab~\cite{Clas}
in order to measure to high accuracy 
 the ratios of the electric (E2) and Coulomb (C2) quadrupole amplitudes to 
the magnetic dipole (M1)
 amplitude.
If both the nucleon and the $\Delta$
are spherical, then E2 and C2 are
expected to be zero. Although 
M1 is indeed the dominant amplitude,
there is mounting experimental evidence over  a range of momentum transfers
that E2 and C2 are
 non-zero~\cite{cnp}.
Similarly in lattice QCD, for hadrons with spins larger than 
1/2, the deformation is determined
by measuring their quadrupole moment
knowing the hadron wave function, which can be obtained 
via density correlators~\cite{ddcorr,cairns}. 
Using these techniques, it was shown  that 
the rho has a non-spherical spatial distribution with a non-zero 
quadrupole moment and that  the
$\Delta$ acquires   
a small deformation as the quark mass decreases~\cite{ddcorr}.
However, direct contact with experiment is established by
 calculating the N to $\Delta$ transition form factors.

In this work we 
calculate these form factors as a function of the momentum transfer
 in lattice QCD in the quenched approximation on a lattice of size
$32^3\times 64$ at $\beta=6.0$.
We obtain, for the first time, accurate results for the E2 and C2
moments for momentum transfer squared, $q^2$,  up to about 1.5~GeV$^2$. 
Our results
are sufficiently accurate to exclude a zero value.
The two novel aspects, as applied to the N to $\Delta$
matrix elements, that are crucial for obtaining 
this accuracy are: 1) An optimal combination of three-point functions,
which allows momentum transfers 
in a spatially symmetric manner
obtained by an appropriate choice of the
interpolating field for the
$\Delta$. 2) An overconstrained analysis using all
lattice momentum vectors contributing to a given $q^2$ value
in the extraction of the 
three transition form factors~\cite{Negele}.

In lattice QCD, transitions 
involving one-photon exchange, such as the one  involved in the 
N to $\Delta$ transition, require the evaluation of
three-point functions. The standard procedure to evaluate a
three-point function
is  to
compute the sequential propagator. This  can be done in two ways:
 In the first approach used in  previous
 lattice calculations~\cite{Leinweber,paper1}, 
the photon couples to a
quark at a fixed time $t_1$  carrying a fixed
momentum ${\bf q}$. This means that 
the form factors can only be evaluated at one
value of the  momentum transfer. Since the current must have a fixed direction and a
fixed momentum this approach is referred to as the fixed current approach. Within this approach
one can use any initial and final state without requiring further inversions, which are
the time consuming part of the evaluation of three-point functions. 
In the second approach, which is used in this work,
we require that the initial state created at time zero has the nucleon quantum numbers  
and  the final state,  annihilated at a fixed time $t_2$, has the $\Delta$ quantum numbers.
The current  
can couple to any time slice $t_1$ carrying any possible value
of  the lattice momentum~\cite{latt03,cairns}. Because the quantum numbers of the final
state are fixed we refer to the second method as  
 the fixed sink method. With the  improvements implemented in
this work, this method  becomes clearly  superior to the fixed current 
approach 
 allowing accurate evaluation of the form factors as a function of $q^2$.

The matrix element for the $\gamma N \> \rightarrow \> \Delta$
transition with on-shell nucleon and $\Delta$ states and real or
virtual photons has the form~\cite{Jones73}
\beq
 \langle \; \Delta (p',s') \; | j_\mu | \; N (p,s) \rangle &=& \nonumber \\  
&\>& \hspace*{-4cm} i   \sqrt{\frac{2}{3}} \biggl(\frac{m_{\Delta}\; m_N}{E_{\Delta}({\bf p}^\prime)\;E_N({\bf p})}\biggr)^{1/2} 
  \bar{u}_\tau (p',s') {\cal O}^{\tau \mu} u(p,s) \;
\label{DjN}
\eeq
where $p(s)$ and $p'(s')$ denote initial and final momenta (spins) and 
$ u_\tau (p',s')$ is a spin-vector in the Rarita-Schwinger formalism.
The operator 
${\cal O}^{\tau \mu}$  can be decomposed in terms of the Sachs form factors
as
\be
{\cal O}^{\tau \mu} =
  {\cal G}_{M1}(q^2) K^{\tau \mu}_{M1} 
+{\cal G}_{E2}(q^2) K^{\tau \mu}_{E2} 
+{\cal G}_{C2}(q^2) K^{\tau \mu}_{C2} \;,
\ee
where the magnetic dipole, ${\cal G}_{M1}$, the electric quadrupole, 
${\cal G}_{E2}$,
 and the Coulomb 
quadrupole, ${\cal G}_{C2}$, form factors depend on the momentum
transfer $q^2 = (p'-p)^2$. The kinematical functions 
$K^{\tau \mu}$ in Euclidean space 
are given in ref.~\cite{paper1}.
Using the relations given in refs.~\cite{Jones73,Gellas}
the ratios  $R_{EM}$ and   $R_{SM}$
in
 the  rest frame of the $\Delta$  
are obtained from the Sachs form factors via 
\be
 R_{EM}= -\frac{{\cal G}_{E2}(q^2)}{{\cal G}_{M1}(q^2)} 
\hspace*{0.5cm}
 R_{SM}=-\frac{|{\bf q}|}{2m_\Delta}\;\frac{{\cal G}_{C2}(q^2)}{{\cal G}_{M1}(q^2)} \quad.
\label{CMR}
\ee
The ratio $R_{EM}$ is also known as EMR and $R_{SM}$ as CMR.

To extract the N to $\Delta$ matrix element 
from lattice measurements we calculate, besides the three 
point function 
$G^{\Delta j^\mu N}_{\sigma} (t_2, t_1 ; {\bf p}^{\;\prime}, {\bf p};\Gamma )$,
the nucleon and $\Delta$ two-point functions, $ G^{NN}$ 
and $ G^{\Delta \Delta}_{ij}$ and look for a plateau in  
the large Euclidean
time behavior of the ratio 
\small
\beq
R_\sigma (t_2, t_1; {\bf p}^{\; \prime}, {\bf p}\; ; \Gamma ; \mu) &=&
\frac{\langle G^{\Delta j^\mu N}_{\sigma} (t_2, t_1 ; {\bf p}^{\;\prime}, {\bf p};\Gamma ) \rangle \;}{\langle G^{\Delta \Delta}_{ii} (t_2, {\bf p}^{\;\prime};\Gamma_4 ) \rangle \;} \> \nonumber \\
&\>& \hspace*{-3.8cm}\biggl [ \frac{ \langle G^{N N}(t_2-t_1, {\bf p};\Gamma_4 ) \rangle \;\langle 
G^{\Delta \Delta}_{ii} (t_1, {\bf p}^{\;\prime};\Gamma_4 ) \rangle \;\langle 
G^{\Delta \Delta}_{ii} (t_2, {\bf p}^{\;\prime};\Gamma_4 ) \rangle \;}
{\langle G^{\Delta \Delta}_{ii} (t_2-t_1, {\bf p}^{\;\prime};\Gamma_4 ) \rangle \;\langle 
G^{N N} (t_1, {\bf p};\Gamma_4 ) \rangle \;\langle 
G^{N N} (t_2, {\bf p};\Gamma_4 ) \rangle \;} \biggr ]^{1/2} \nonumber \\
&\;&\hspace*{-1cm}\stackrel{t_2 -t_1 \gg 1, t_1 \gg 1}{\Rightarrow}
\Pi_{\sigma}({\bf p}^{\; \prime}, {\bf p}\; ; \Gamma ; \mu) \; .
\label{R-ratio}
\eeq
\normalsize
We use the lattice conserved   electromagnetic current,   $j^\mu (x)$,
symmetrized on site $x$ by taking
$
j^\mu (x) \rightarrow \left[ j^\mu (x) + j^\mu (x - \hat \mu) \right]/ 2
$
and projection matrices for the Dirac indices
\be
\Gamma_i = \frac{1}{2}
\left(\begin{array}{cc} \sigma_i & 0 \\ 0 & 0 \end{array}
\right) \;\;, \;\;\;\;
\Gamma_4 = \frac{1}{2}
\left(\begin{array}{cc} I & 0 \\ 0 & 0 \end{array}
\right) \;\; .
\ee
Throughout this work we use kinematics where the $\Delta$ is produced at rest
and therefore ${\bf q}={\bf p}^{\prime}-{\bf p}=-{\bf p}$.
We fix $t_2=12$ in lattice units  and search for a plateau of 
 $ R_{\sigma}(t_2,t_1;{\bf p}^{\; \prime}, {\bf p}\; ; \Gamma ;\mu)$ as a function of $t_1$.  $Q^2=-q^2$ denotes the Euclidean momentum transfer squared.

We can extract the three Sachs form factors from the ratio of Eq.~(\ref{R-ratio}) by choosing
appropriate combinations of $\sigma$ indices  and $\Gamma$ matrices.
However there are several choices of $\sigma$ indices and $\Gamma$
matrices that can be used each requiring an inversion. Therefore we must determine the
most suitable combination of three-point functions from which to extract
the Sachs form factors. 

For example the dipole form factor can be extracted
from  
\be 
\Pi_{\sigma}({\bf q}\; ; \Gamma_4 ;\mu)= i A \epsilon^{\sigma 4\mu j} p^j {\cal G}_{M1}(Q^2) \quad ,
\label{pure GM1}
\ee
where A is a kinematical coefficient. This
 means that there are six statistically independent matrix elements
to extract ${\cal G}_{M1}$
each requiring the evaluation of a sequential propagator. 
However, due to the epsilon factor,
a choice of one of the six 
combinations means
that only momentum transfers in  one direction contribute. Instead
if we take the symmetric combination,
\be
S_1({\bf q};\mu)=  \sum_{\sigma=1}^3\Pi_\sigma({\bf q}\; ; \Gamma_4 ;\mu) 
 \quad ,
\label{S1}
\ee
lattice momentum vectors in all directions contribute. This combination, which
we refer to as sink type $S_1$, is built into the $\Delta$ interpolating
field and  requires only one inversion.

Another choice of three-point functions is to use the projection matrices
$\Gamma_k$ instead of $\Gamma_4$. 
%
%
The relations are more involved in this case and will
be discussed in detail in a forthcoming publication~\cite{futurepaper}. However,
as in the example given in Eq.~(\ref{pure GM1}), for any current direction, 
one can choose any of the
six statistically different three-point functions from which to extract the Sachs
form factors. Instead  of choosing one of  six we can consider a linear combination
that involves, in a symmetric manner, all spatial directions
allowing,  for a given $Q^2$, the maximum number of momentum vectors to contribute. We take
\be
S_2({\bf q};\mu)=\sum_{\sigma\neq k=1}^{3} \Pi_\sigma({\bf q}\; ; \Gamma_k ;\mu) \quad, 
\label{S2}
\ee
 which we refer to as sink $S_2$. 
When the current is in the spatial direction 
both ${\cal G}_{E2}$ and
${\cal G}_{C2}$ can be extracted from $S_2$ with one inversion.
 In addition, when the current is in the time direction, 
 $S_2$ provides a statistically independent way for 
evaluating
 $  {\cal G}_{C2}$, with no extra
cost.
 Another combination to extract  ${\cal G}_{E2}$ and  ${\cal G}_{C2}$
is $S_3({\bf q};\mu)= \Pi_3({\bf q}; \Gamma_3 ;\mu)
-\bigl(\Pi_1({\bf q}; \Gamma_1 ;\mu)+\Pi_2({\bf q}; \Gamma_2 ;\mu)\bigr)/2$,
 which produces results
of comparable quality to those obtained with $S_2$~\cite{latt04}. 
In the case of E2, sink type $S_3$ has the disadvantage of vanishing  at the lowest value of $Q^2$ 
 whereas $S_2$ contributes at all values of $Q^2$. 
For C2, on the other hand, $S_2$ gives zero at the lowest $Q^2$, whereas
source type $S_3$ gives a non-vanishing result.

The second important ingredient in the extraction of the form factors
 is to take into
account in our analysis  all the lattice momentum vectors that contribute to a given 
$Q^2$. This is done by solving the overcomplete set of equations
\be
P({\bf q};\mu)= D({\bf q};\mu)\cdot F(Q^2) 
\ee
where $P({\bf q};\mu)$ are the lattice measurements of the ratio
given in Eq.~(\ref{R-ratio}) having statistical errors
$w_k$ and using the different sink types,
$F =  \left(\begin{array}{c} {\cal G}_{M1} \\
                                   {\cal G}_{E2} \\ 
                                   {\cal G}_{C2} \end{array}\right)$
and, with $N$ being the number of current 
directions and momentum vectors contributing to 
a given $Q^2$,  D is an $N\times 3$ matrix which depends on 
kinematical factors. We extract the form factors by 
minimizing 
\be
\chi^2=\sum_{k=1}^{N} \Biggl(\frac{\sum_{j=1}^3 D_{kj}F_j-P_k}{w_k}\Biggr)^2
\ee
using the singular value decomposition of D.

\begin{figure}[h]
\epsfxsize=8.5truecm
\epsfysize=5.5truecm
\mbox{\epsfbox{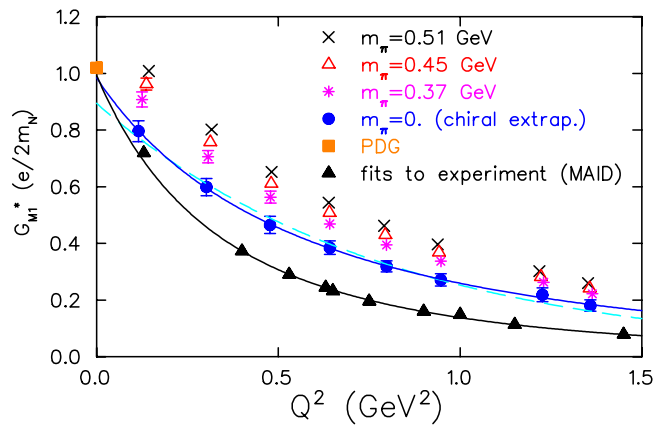}}
\caption{${\cal G}_{M1}^*$ as function of $Q^2$. Results at $\kappa=0.1554$
are shown by the crosses, at $\kappa=0.1558$ by the open triangles
and at $\kappa=0.1562$ by the asterisks. The filled circles 
are  results in the chiral limit and the filled triangles are results
 extracted from
fits to the measured
differential cross sections using MAID~\cite{Tiator}. The filled square
is the Particle Data Group result~\cite{PDG}. The solid
lines are fits using the Ansatz of Eq.~(\ref{fit1}).
The dashed line is a fit to the lattice data using the Ansatz $a\exp(-bQ^2)$.}
\label{fig:GM1}
\end{figure}

\begin{figure}[h]
\epsfxsize=8.5truecm
\epsfysize=8.5truecm
\mbox{\epsfbox{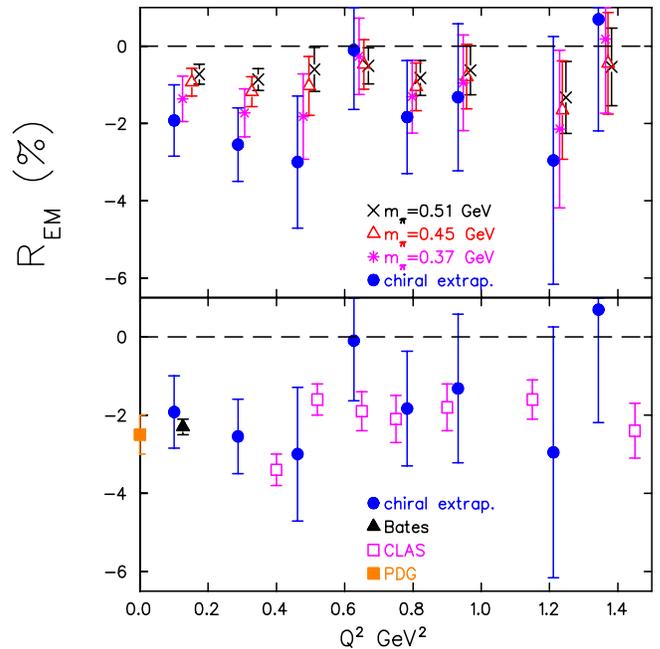}}
\caption{$R_{EM}$ as a function of $Q^2$.  Upper graph
shows results at
$\kappa=0.1554$ (crosses), at $\kappa=0.1558$ (open triangles) 
and at $\kappa=0.1562$ ( asterisks). The filled circles show
chiral limit extrapolations. Lower graph shows The filled triangles show 
recent experimental results~\cite{Bates} (filled triangles) and
 \cite{Clas} )opensqures). The filled square
is the Particle Data group result~\cite{PDG}.}
\label{fig:EMR}
\vspace*{-0.5cm}
\end{figure}

\begin{figure}
\epsfxsize=8.5truecm
\epsfysize=8.5truecm
\mbox{\epsfbox{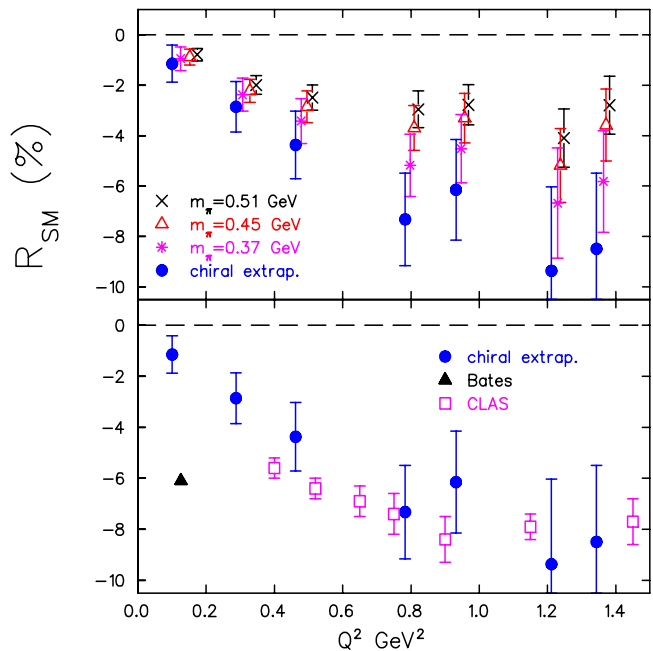}}
\caption{$R_{SM}$ as a function of $Q^2$.  
The notation is the same as that of Fig.~\ref{fig:EMR}}
\label{fig:CMR}
\end{figure}

All the  results for the form factors are obtained 
using 200 configurations and three values of 
the  hopping parameter $\kappa$. The values
of $\kappa$ chosen are
$0.1554, 0.1558$ and 0.1562 and give ratio of pion to rho mass
$m_\pi/m_\rho=0.64, 0.59$ and $0.50$ respectively. We use the nucleon
mass at the chiral limit to set the lattice spacing $a$ obtaining
$a^{-1}=2.04(2)$~GeV (a=0.098~fm).
Using the optimized sink $S_1$ we 
show in Fig.~\ref{fig:GM1} at the  three quark masses 
our results for the Ash form factor ${\cal G}_{M1}^*$
defined by~\cite{Ash} 
\be 
{\cal G}_{M1}^*(Q^2) = 
\frac{1}{3}\>\frac{1}{\sqrt{1+\frac{Q^2}{(m_N+m_\Delta)^2}}}
 \>{\cal G}_{M1}(Q^2) \quad.
\label{GM1*}
\ee
To obtain the results at the chiral
limit, shown on the same figure, we perform a linear extrapolation in  
 $m_\pi^2$. Although we expect chiral logs
that appear at next-to-leading order in chiral perturbation theory to be
suppressed for the momentum transfers studied in this work, 
our linear extrapolation introduces a systematic uncertainty. This uncertainty
can not be assessed, since  the known chiral perturbation theory results~\cite{Gellas, Arndt}
are valid at very low masses and momentum transfers.
On the same figure we also show the 
experimental values as extracted from the measured cross sections using the 
phenomenological model MAID~\cite{Tiator}. We 
perform fits to both the lattice data at the chiral
limit and to the MAID data using
the phenomenological parametrization
\be
{\cal G}_{a}(Q^2) = {\cal G}_a(0) R_a (Q^2) G_E^p(Q^2)
\label{fit1}
\ee
where $R_a (Q^2)$ for $a=M1, E2$ and $C2$ measures the deviations 
from the proton electric form factor $ G_E^p(Q^2)=1/(1+Q^2/0.71)^2$. 
Usually experimental data are fitted by taking
$
 R_{M1} (Q^2) =  R_{E2} (Q^2) =  R_{C2} (Q^2)  = 1 + \alpha \exp(-\gamma Q^2)$
~\cite{Sato}.
As can be seen this fitting Ansatz provides a very good description
to the MAID data.
Although the lattice data at the chiral limit lie higher than the MAID data
they
can be fitted to the same form yielding at $Q^2=0$ a value
consistent with that given by the Particle Data Group~\cite{PDG}.
The lattice data are also well described by the simple exponential Ansatz 
$a \exp(-bQ^2)$, which however at $Q^2=0$ gives a value lower than
experiment.

Using the optimized sink $S_2$ we 
extract the quadrupole form factors,
${\cal G}_{E2}$ and
${\cal G}_{C2}$, at three values of the quark mass,
as shown in 
 Figs.~\ref{fig:EMR} and~\ref{fig:CMR}. 
We note that ${\cal G}_{C2}$ at the lowest $Q^2$ value is extracted
using source type $S_3$ since $S_2$ vanishes for this particular 
lattice vector.
On the same figures
we also show the values obtained
in the chiral limit 
by performing a  linear extrapolation in $m_\pi^2$. As expected, 
both EMR and
CMR become more negative as we approach the chiral limit. 
Our results for EMR and CMR at the chiral limit are compared
to recent measurements~\cite{Bates,Clas} in Figs.~\ref{fig:EMR} 
and~\ref{fig:CMR}, respectively. 
The quenched results for EMR are accurate enough
to exclude a zero value  at low
$Q^2$. With our current statistics they are in agreement
with the experimental measurements.
Whether the apparent discrepancy for CMR at low $Q^2$ is a significant 
deficiency of quenched QCD or a problem with a single data point remains 
to be resolved by new measurements that are currently being analyzed. 
As $Q^2$ increases, however,  the quenched
results agree qualitatively with measurements.



In summary, two novel methods are applied in the
 evaluation
of the N to $\Delta$ transition form factors: The first improvement
comes from employing an optimized sink for the $\Delta$ allowing
a maximum number of lattice matrix elements to contribute and the  
second 
 from utilizing this enlarged set of data in
an overconstrained analysis.
Given that  there are  ambiguities 
in the extraction of the quadrupole amplitudes from  experimentally 
measured  response functions arising from  using models, 
accurate lattice data are extremely valuable:
excluding a zero quadrupole stength in lattice QCD corroborates
experimental observations for  a non-zero $R_{EM}$ and $R_{SM}$.
This is particularly important for  CMR where at low $Q^2$ there are
very few  accurate experimental measurements.
If confirmed, the agreement of EMR with experiment at low $Q^2$ while CMR disagrees with experiment 
raises interesting questions regarding the pion cloud
contributions to these ratios  due to the absence of the sea quarks.
Having, for the first time, demonstrated that we
can reliably extract CMR in quenched lattice QCD opens  the way for a precise investigation
of sea quark contributions to both EMR and CMR, which can lead
to an understanding of the physical mechanism
responsible for non-zero quadrupole strength in the N to $\Delta$ transition.

{\bf Acknowledgments:} 
A.T. is supported by the Levendis Foundation and W.S. is partially supported by the Alexander von Humboldt Foundation .

  This research used resources of the National Energy Research Scientific 
Computing Center, which is supported by the Office of Science of the U.S. 
Department of Energy under Contract No. DE-AC03-76SF00098. This work is 
supported in part by the U.S. Department of Energy (D.O.E.) under 
cooperative research agreement $\#$ DF-FC02-94ER40818 and
$\#$ DE-FC02-01ER41180.


\begin{thebibliography}{99}
\bibitem{Bates} C.Mertz {\it et al.}, Phys. Rev. Lett. {\bf 86}, 2963 (2001).
\bibitem{Clas} K. Joo {\it et al.}, 
Phys.\ Rev.\ Lett.\ {\bf 88} 122001 (2002).
\bibitem{cnp} C. N. Papanicolas, Eur. Phys. J. A18,  141 (2003); A. M. Bernstein, Eur. Phys. J. A17,  349 (2003).
\bibitem{ddcorr}C. Alexandrou, Ph. de Forcrand and A. Tsapalis, Phys. Rev. D
{\bf 66}, 094503 (2002); Phys. Rev. D {\bf 68}, 074504 (2003); Nucl.Phys. (Proc. Suppl.) {\bf 119}, 422 (2003); Nucl. Phys. {\bf A721}, 907 (2003); 
Nucl. Phys. (Proc.Suppl.) {\bf 129}, 221 (2004).
\bibitem{cairns} C. Alexandrou, Nucl. Phys. (Proc.Suppl.) {\bf 128}, 1 (2004).
\bibitem{Negele} LHPC and SESAM collaboration, Ph. Hagler, {\it et. al}, Phys. Rev. D {\bf 68}, 034505 (2003). 
\bibitem{Leinweber} D. B. Leinweber, T. Draper, and 
R. M. Woloshyn,
Phys. Rev. D {\bf 48}, 2230 (1993).
\bibitem{paper1} C. Alexandrou, Ph. de Forcrand, Th. Lippert, H. Neff, J.W. Negele, K. Schilling, W. Schroers and A. Tsapalis, Phys. Rev. D {\bf 69}, 114506 (2004).
\bibitem{latt03} C. Alexandrou, {\it et. al},
 Nucl. Phys. (Proc.Suppl.) {\bf 129}, 302 (2004).
\bibitem{Jones73} H. F. Jones and M.C. Scadron, Ann. Phys. (N.Y.) {\bf81},
1 (1973).
\bibitem{Gellas} G. C. Gellas, T. R. Hemmert, 
C. N. Ktorides, and 
G. I. Poulis, Phys. Rev. D {\bf 60}, 054022 (1999).
\bibitem{futurepaper} C. Alexandrou, {\it et al.}, in preparation.
\bibitem{latt04} C. Alexandrou, {\it et al.}, hep-lat/0408017.
\bibitem{Tiator} D. Drechsel, O. Hanstein, S.S. Kamalov and L. Tiator, 
 Nucl. Phys. {\bf A645} 145 (1999); L. Tiator, {\it et. al}
 Nucl.Phys. {\bf A689}, 205 (2001).
\bibitem{PDG} K. Hagiwara, {\it et al.}, Phys. Rev. D {\bf 66}, 010001 (2002).
\bibitem{Ash} W. W. Ash,
K. Berkelman, C. A. Lichtenstein, A. Ramanauskas and 
R. H. Siemann, Phys. Lett. {\bf B24}, 165 (1967).
\bibitem{Arndt}  D. Arndt and B. C. Tiburzi, Phys. Rev. D {\bf 69}, 014501 
(2004). 
\bibitem{Sato} T. Sato and T.-S. H. Lee, Phys. Rev., C{\bf 63},  055201 (2001).
\end{thebibliography}
\end{document}